\documentclass[]{gHPR2e}
\usepackage{graphicx}
\usepackage{amssymb}
\usepackage{amsmath}

\begin{document}
\doi{10.1080/0895795YYxxxxxxxx}
\issn{1477-2299}
\issnp{0895-7959} \jvol{00} \jnum{00} \jyear{2010} \jmonth{March}

\newcommand{\bk}{{\bf k}}
\newcommand{\bq}{{\bf q}}
\newcommand{\bQ}{{\bf Q}}
\newcommand{\bG}{{\bf G}}
\newcommand{\bK}{{\bf K}}
\newcommand{\bp}{{\bf p}}
\newcommand{\bx}{{\bf x}}
\newcommand{\by}{{\bf y}}
\newcommand{\br}{{\bf r}}
\newcommand{\bR}{{\bf R}}
\newcommand{\bJ}{{\bf J}}
\newcommand{\bz}{{\bf 0}}
\newcommand{\ba}{{\bf a}}
\newcommand{\bh}{{\bf h}}
\newcommand{\bd}{{\bf d}}
\newcommand{\bv}{{\bf v}}
\newcommand{\bdelta}{{\boldsymbol\delta}}
\newcommand{\Li}{{\mathop{\rm{Li}}\nolimits}}
\newcommand{\cotg}{{\mathop{\rm{cotg}}\nolimits}}
\renewcommand{\Im}{{\mathop{\rm{Im}}\nolimits\,}}
\renewcommand{\Re}{{\mathop{\rm{Re}}\nolimits\,}}
\newcommand{\sgn}{{\mathop{\rm{sgn}}\nolimits\,}}
\newcommand{\Tr}{{\mathop{\rm{Tr}}\nolimits\,}}
\newcommand{\EF}{E_{\mathrm{F}}}
\newcommand{\kB}{k_{\mathrm{B}}}
\newcommand{\kF}{k_{\mathrm{F}}}
\newcommand{\nF}{n_{\mathrm{F}}}
\newcommand{\Green}{{G}}
\newcommand{\dSC}{{\mathrm{dSC}}}
\newcommand{\dPG}{{\mathrm{dPG}}}
\newcommand{\dDW}{{\mathrm{dDW}}}
\newcommand{\Ret}{{\mathrm{R}}}
\newcommand{\Tau}{T_\tau}
\newcommand{\modified}[1]{{\relax #1}}

\markboth{Pellegrino, Angilella, Pucci}{Strain effect in graphene}


\title{Effect of uniaxial strain on the reflectivity of graphene}

\author{F. M. D. Pellegrino$^{\rm a,b,c}$,
G. G. N. Angilella$^{\rm a,b,c,d}$$^{\ast}$\thanks{$^\ast$Corresponding author.
Email: giuseppe.angilella@ct.infn.it\vspace{6pt}}
and R. Pucci$^{\rm a,d}$\\\vspace{6pt}  
$^{\rm a}${\em{Dipartimento di Fisica e Astronomia, Universit\`a di Catania, Via
S. Sofia, 64, I-95123 Catania, Italy}}; $^{\rm b}${\em{Scuola Superiore di
Catania, Via S. Nullo, 5/i, I-95123 Catania, Italy}}; $^{\rm c}${\em{INFN, Sez.
Catania, I-95123 Catania, Italy}}; $^{\rm d}${\em{CNISM, UdR Catania, I-95123
Catania, Italy}}\\\vspace{6pt}\received{\today}}

\maketitle

\begin{abstract}

We evaluate the optical reflectivity for a uniaxially strained graphene single
layer between a SiO$_2$ substrate and air. A tight binding model for the band
dispersion of graphene is employed. As a function of the strain modulus and
direction, graphene may traverse one of several electronic topological
transitions, characterized by a change of topology of its Fermi line. This
results in features in the conductivity within the optical range, which might be
observable experimentally.\bigskip

\bigskip
\end{abstract}

\section{Introduction}

Graphene is an atomically thick single layer of carbon atoms in the $sp^2$
hybridization status, linked to form a honeycomb lattice. This almost unique
two-dimensional (2D) structure may be thought of as the building block of
three-dimensional (3D) graphite, one-dimensional (1D) nanotubes, and
zero-dimensional (0D) fullerenes. The electronic properties of graphene have
been known since decades \cite{Wallace:47}. The band structure consists of two
bands, touching at the Fermi level in a linear, cone-like fashion at the
so-called Dirac points $\pm\bK$. Indeed, the low-energy electronic properties of
graphene can be mapped onto those of relativistic massless particles, thus
allowing the observation in a solid state system of several effects predicted by
quantum electrodynamics \cite{CastroNeto:08}.

Probably less well known are graphene's equally outstanding mechanical
properties. Let alone the very existence of such a 2D crystal, challenging
Mermin-Wagner's theorem \cite{CastroNeto:08} through the development of
relatively small ripples which slightly perturb planarity, graphene achieves the
record breaking strength of $\sim 40$~N/m \cite{Lee:08}, which nearly equals the
theoretical limit, whereas its Young modulus of $\sim 1.0$~TPa \cite{Lee:08}
places graphene amongst the hardest materials known. Its electronic, optical,
and elastic properties can be strongly modified by uniaxial strain
\cite{Pereira:08a} as well as local distortions, such as those caused by local
impurities \cite{Pellegrino:09}. In particular, recent \emph{ab initio}
calculations \cite{Liu:07} as well as experiments \cite{Kim:09} have
demonstrated that graphene single layers can reversibly sustain elastic
deformations as large as 20\%. In this context, it has been shown that Raman
spectroscopy can be used as a sensitive tool to determine the strain as well as
some strain-induced modifications of the electronic and transport properties of
graphene \cite{Ni:08,Mohiuddin:09}.

Here, we will be concerned on the effects induced by applied strain on the
optical reflectivity of graphene. Based on our previous results for the strain
effect on the optical conductivity of graphene \cite{Pellegrino:09b}, we will
discuss the frequency dependence of the reflectivity, as a function of strain
modulus and direction. We will interpret our results in terms of the proximity
to several electronic topological transitions (ETT).

\section{Model}

The tight-binding Hamiltonian for a honeycomb lattice can be written as
\begin{equation}
H = \sum_{\bR,\ell} t_\ell a^\dag (\bR) b(\bR+\bdelta_\ell) + \mathrm{H.c.},
\label{eq:H}
\end{equation}
where $a^\dag (\bR)$ is a creation operator on the position $\bR$ of the A
sublattice, $b(\bR+\bdelta_\ell)$ is a destruction operator on a nearest
neighbor (NN) site $\bR+\bdelta_\ell$, belonging to the B sublattice, and 
$\bdelta_\ell$ are the vectors connecting a given site to its NNs, their
unstrained values being $\bdelta_1^{(0)} = a(1,\sqrt{3})/2$,  $\bdelta_2^{(0)} =
a(1,-\sqrt{3})/2$,  $\bdelta_3^{(0)} = a(-1,0)$, with $a=1.42$~\AA, the
equilibrium C--C distance in a graphene sheet \cite{CastroNeto:08}. In
Eq.~(\ref{eq:H}), $t_\ell \equiv t(\bdelta_\ell )$, $\ell=1,2,3$, is the hopping
parameter between two NN sites. In the absence of strain they reduce to a single
constant, $t_\ell \equiv t_0$, with $t_0 = -2.8$~eV \cite{Reich:02}. 

In terms of the strain tensor \cite{Pereira:08a}
\begin{equation}
{\boldsymbol\varepsilon} = \varepsilon
\begin{pmatrix}
\cos^2 \theta -\nu \sin^2 \theta & (1+\nu)\cos\theta\sin\theta \\
(1+\nu)\cos\theta\sin\theta & \sin^2 \theta -\nu \cos^2 \theta
\end{pmatrix} ,
\label{eq:strainmat}
\end{equation}
the deformed lattice distances are related to the relaxed ones by $\bdelta_\ell
= (\mathbb{I} +  {\boldsymbol\varepsilon}) \cdot \bdelta^{(0)}_\ell$. In
Eq.~(\ref{eq:strainmat}), $\theta$ denotes the angle along which the strain is
applied, with respect to the $x$ axis in the lattice coordinate system,
$\varepsilon$ is the strain modulus, and $\nu=0.14$ is Poisson's ratio, as
determined from \emph{ab initio} calculations for graphene \cite{Farjam:09}, to
be compared with the known experimental value $\nu=0.165$ for graphite
\cite{Blakslee:70}. The special values $\theta=0$ and $\theta=\pi/6$ refer to
strain along the armchair and zig~zag directions, respectively.

The band structure derived from Eq.~(\ref{eq:H}), also including overlap between
NNs \cite{Pellegrino:09}, is characterized by two bands $E_{\bk\lambda}$
($\lambda=1,2$) touching cone-like at the Fermi level at the Dirac points
$\pm\bk_D$ \cite{Pellegrino:09b}. While at zero strain these coincide with the
high symmetry points $\pm\bK$ at the vertices of the hexagonal first Brillouin
zone (1BZ) of graphene, their positions move away from $\pm\bK$ with increasing
strain. Depending on the strain modulus and strength, with increasing strain
they may eventually merge into the midpoints of one of the sides of the 1BZ. In
that case, the cone approximation breaks down, and a finite gap opens at the
Fermi level \cite{Pellegrino:09b}. For intermediate strains, the constant energy
contours of $E_{\bk\lambda}$ can be grouped according to their topology, and are
divided by three separatrix lines \cite{Pellegrino:09b}. This corresponds to
having three distinct electronic topological transitions (ETT)
\cite{Lifshitz:60,Blanter:94,Varlamov:99}, which are here tuned by uniaxial
strain, as has been suggested to be the case in other quasi-2D materials, such
as the cuprates \cite{Angilella:01,Angilella:02d} and the Bechgaard salts
\cite{Angilella:02}. Correspondingly, while the density of states (DOS) is
characterized by a linear behaviour close to the Fermi level, because of the
linearity in $E_{\bk\lambda}$ at $\bk=\pm\bk_D$, a detailed analysis beyond the
cone approximation reveals the presence of logarithmic singularities in the DOS
at higher energies, that may be connected with the proximity to the ETTs
\cite{Pellegrino:09b}. This behaviour is reproduced in the frequency dependence
of the longitudinal optical conductivity $\sigma(\omega)$, which has been
studied at increasing strain modulus and different strain directions
\cite{Pellegrino:09b}.

\begin{figure}[t]
\centering
\includegraphics[height=0.7\columnwidth,angle=-90]{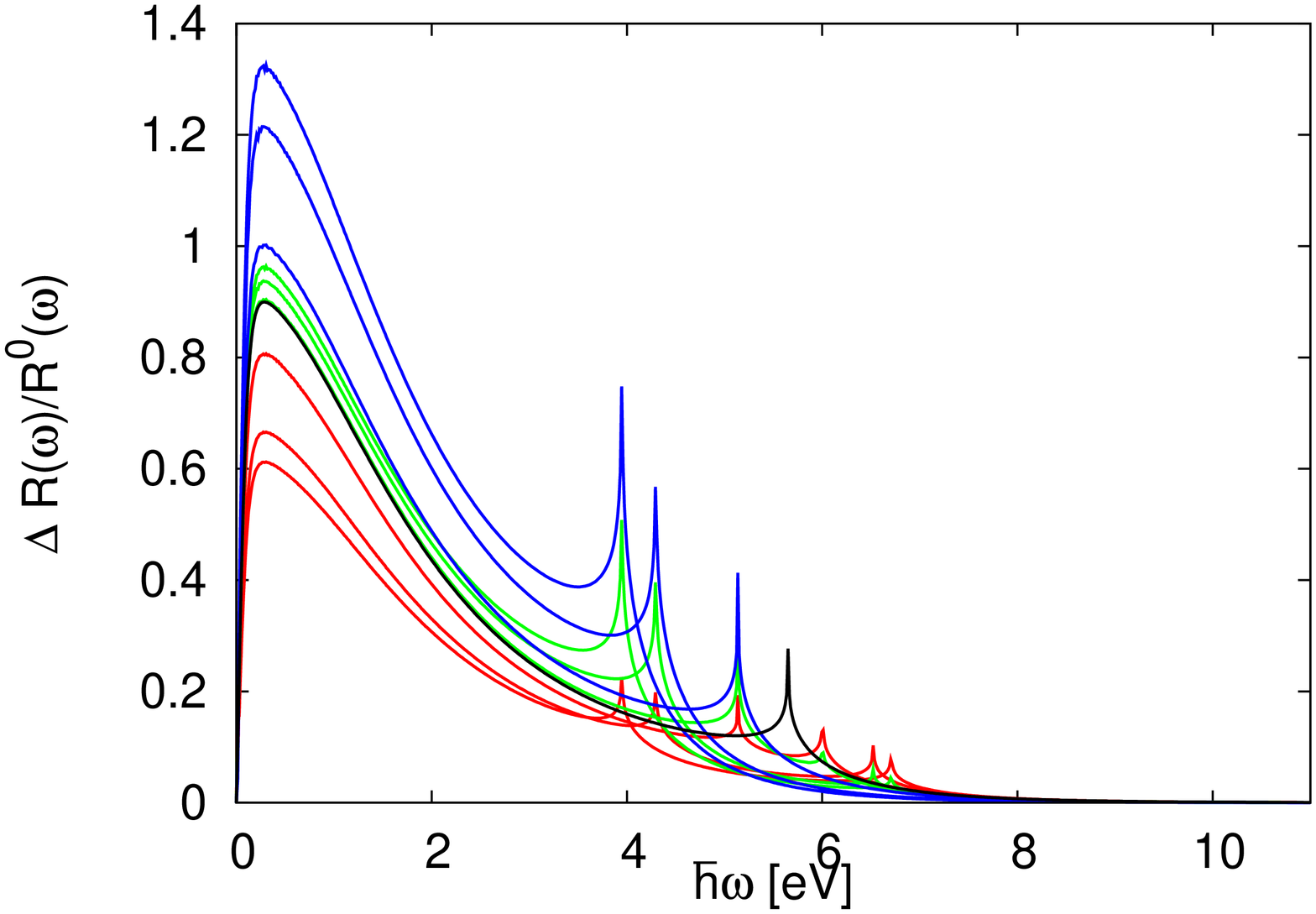}
\includegraphics[height=0.7\columnwidth,angle=-90]{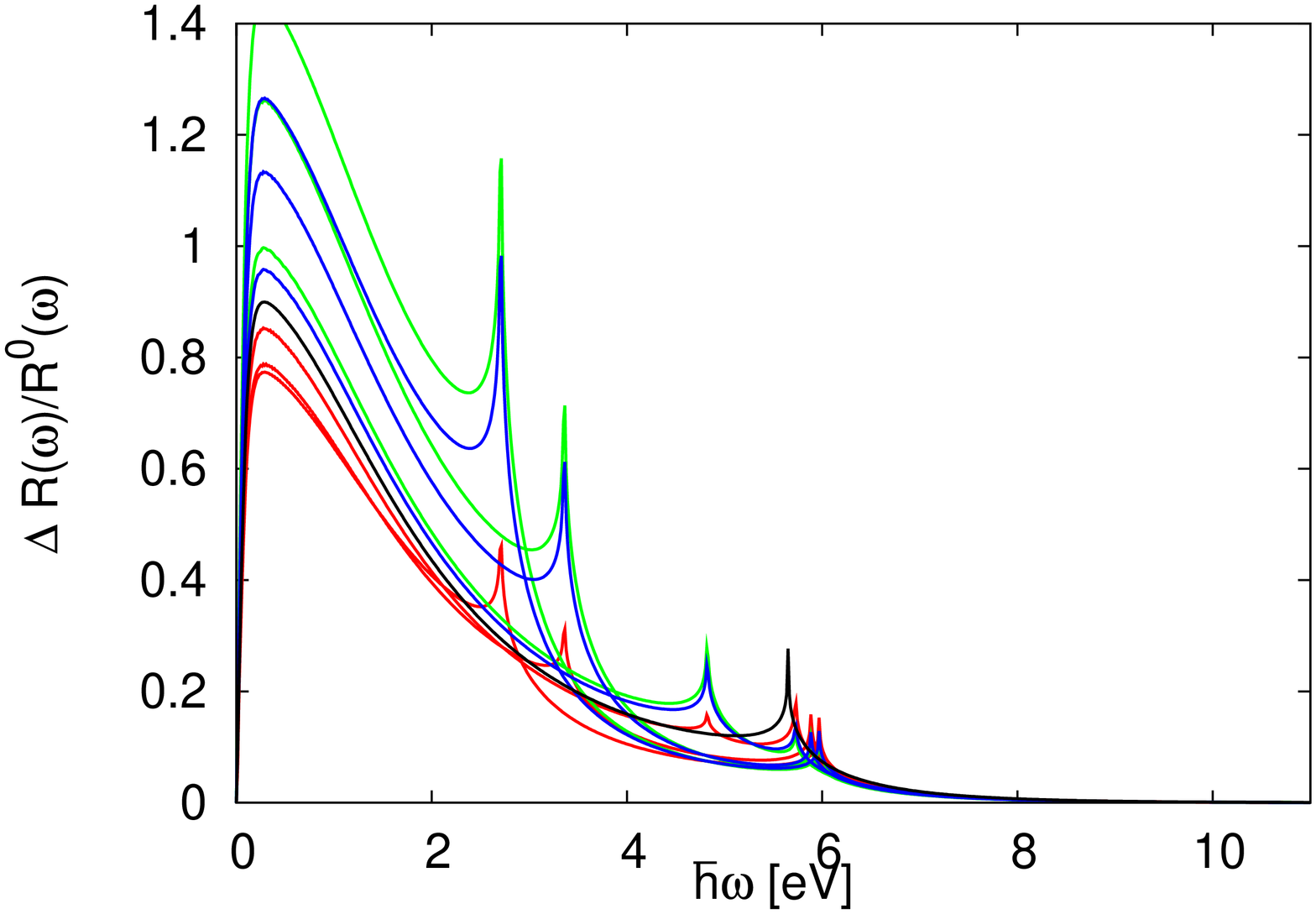}
\caption{(Color online) Relative reflectivity $\Delta R(\omega)/R^0 (\omega)$ as
a function of frequency $\omega$ for a graphene monolayer between SiO$_2$ and
air. Top panel shows cases with strain applied along the $\theta=0$ (armchair)
direction, while bottom panel shows cases with $\theta=\pi/6$ (zig~zag).
Different colors refer to a different polarization of the electric field:
$\phi=0$ (red), $\phi=\pi/4$ (green), $\phi=\pi/2$ (blue). Black solid line
refers to the unstrained case. Within each group of lines, modulus of
applied strain increases as $\varepsilon=0.025$, 0.075, 0.1 (bottom to top).}
\label{fig:reflectivity}
\end{figure}

\section{Results and conclusions}

We consider light scattering across two media, with refraction index $n_i
(\omega)$ ($i=1,2$), separated by a graphene monolayer. In the case of normal
incidence, the reflectivity of such a system can be written as
\cite{Stauber:08a}
\begin{equation}
R(\omega) = \left(\frac{n_1(\omega) + \frac{\sigma(\omega)}{\epsilon_0 c} -n_2
(\omega)}{n_1(\omega) + \frac{\sigma(\omega)}{\epsilon_0 c} +n_2
(\omega)}\right)^2 .
\label{eq:reflectivity}
\end{equation}
Assuming air on top of graphene, we have $n_2 (\omega)=1$, whereas we can model
the frequency dependence of the refraction index of the substrate through
Cauchy's law, as $n_1 (\omega) = n_1 (0) + \omega^2 / \omega_b^2$. In the case
of SiO$_2$, from a fit of data in Ref.~\cite{Jung:07}, we find $n_1 (0) =1.448$
and $E_b = \hbar\omega_b = 3.305$~eV. We evaluate $\sigma(\omega)$ as in
Ref.~\cite{Pellegrino:09b}. 

Fig.~\ref{fig:reflectivity} then shows our results for the relative reflectivity
$\Delta R(\omega)/R^0 (\omega) \equiv [R(\omega)-R^0 (\omega)]/R^0 (\omega)$,
where $R(\omega)$ is the reflectivity of a single graphene monolayer with
applied strain, between SiO$_2$ and air, Eq.~(\ref{eq:reflectivity}), and $R^0
(\omega)$ refers to the case without the insertion of a graphene layer. We
specifically consider applied strain along the armchair ($\theta=0$) and zig~zag
($\theta=\pi/6$) directions. Different colors refer to different polarization
angles of the electric field (though always with normal incidence). In all
cases, the black solid line refers to the case without strain, where
the reflectivity does not depend on the electric field polarization angle.

While at large frequencies the reflectivity invariably tends to the case without
graphene, differences are to be seen in the low-frequency range. In particular,
the presence of logarithmic spikes can be associated with the proximity to the
electronic topological transitions referred to above, their actual position
depending on the strain modulus $\varepsilon$, as well as on the relative
orientation of strain $\theta$ and electric field polarization angle $\phi$.

\bibliographystyle{gHPR}
\bibliography{a,b,c,d,e,f,g,h,i,j,k,l,m,n,o,p,q,r,s,t,u,v,w,x,y,z,zzproceedings,Angilella}
\end{document}